# A beamline for time-resolved extreme ultraviolet and soft x-ray spectroscopy

Jakob Grilj,[1,2] Emily Sistrunk,[1] Markus Koch,[1,3] Markus Gühr [1,a)]

1 Stanford PULSE Institute, SLAC National Accelerator Laboratory, Menlo Park, CA 94025, USA

2 Laboratoire de Spectroscopie Ultrarapide, Ecole Polytechnique Fédérale de Lausanne EPFL, 1015, Switzerland

3 Institute of Experimental Physics, Graz University of Technology, A-8010 Graz, Austria

a) author to whom correspondence should be addressed: mguehr@slac.stanford.edu

## Abstract

High harmonic generation is a convenient way to obtain extreme ultraviolet light from table-top laser systems and the experimental tools to exploit this simple and powerful light source for time-resolved spectroscopy are being developed by several groups. For these applications, brightness and stability of the high harmonic generation is a key feature. This article focuses on practical aspects in the generation of extreme ultraviolet pulses with ultrafast commercial lasers, namely generation parameters and online monitoring as well as analysis of generation yield and stability.

## Introduction

The interaction of light with the valence electrons of molecules and materials presents a central problem for the understanding of life and many technological applications. Vision and light harvesting involve the transfer of light energy into chemical energy,[1–3] the photoprotection of DNA and its components involves the selective transfer of light energy into heat instead of chemical change.[4,5] Materials that change their properties upon light illumination[6–8] as well as photovoltaic systems[9] promise major technological breakthroughs. The photoexcited state of matter is generally described by a complex concerted motion of electrons and nuclei on the timescale of femto– to picoseconds, whose understanding is still challenging[10] demanding extensive experimental studies. The invention of ultrafast lasers has opened this timescale for direct investigations.[11]

Due to the recent progress in strong field laser interactions, femto– and attosecond light pulses in the extreme ultraviolet (EUV) spectral range between 10 and 100 eV photon energy are becoming more widespread in ultrafast laser laboratories. Similar to x-rays, the photons in the EUV spectral domain interact preferentially with atomic core levels, which are genuinely distinct for different chemical elements. Together with the relatively tight spatial extend of the corresponding core wavefunction, the EUV pulses are used for spatially and element selective spectroscopy of matter. The most important core levels in the EUV spectral domain are the 3p and 4p states of the 3d and 4d transition metals.[12] Pioneering time resolved studies using EUV probe pulses were performed in atoms and molecules[13,14] as well as on materials [15–19] and have recently also been extended to solvated systems.[20]

The generation mechanism for the EUV pulses is based on the sub-optical cycle interaction of a strong infrared (IR) laser field with a rare gas target, which is generally called high harmonic generation (HHG). During one optical half-cycle, part of the valence electrons are ionized from an atom. The electron gains kinetic energy in the IR field which can be transferred into photon energy upon inelastic collision of the electron with the remaining ion.[21–23] The small cross section for this single atom inelastic collision process can only partially be compensated by coherent phase matching over the excitation medium. Unlike light from other nonlinear processes, the HHG-EUV light carries an intensity dependent phase, making phase matching much more demanding.[24–26] Popular schemes to make use of some phase matching include gas jets,[27–29] gas cells,[16,30–34] multiple gas cells in series,[35] hollow core capillaries,[36] and plasma plumes.[37]

Femotosecond EUV and soft x-ray (SXR) light pulses can also be generated in synchrotrons in a so-called slicing beam line[38] which can deliver sub-picosecond time resolution though at largely reduced flux compared to normal operation. Despite the experimental challenges, they have been successfully used to elucidate photochemical reactions.[39] The free electron lasers (FEL) FLASH and Fermi@Elettra can generate femtosecond EUV light pulses. Fermi@Elettra is seeded by an external laser with UV light but seeding with shorter wavelength pulses from a HHG source is envisaged[40] and has been successfully demonstrated in a test facility at Spring 8.[41]

In this paper, we present a beamline for optimal transmission of the EUV light while filtering the strong infrared field. We present a simple yet effective online monitor for the relative intensity of the EUV light, which allows us to optimize and monitor the HHG efficiency. The device is based on a grazing incidence mirror for the EUV pulses. We present our approach of optimizing the HHG process in a gas cell and present systematic variations with cell length, target gas pressure and position with respect to the IR driver pulse. The beamline can be conveniently coupled to an optical excitation pulse for realizing time resolved absorption, photoemission and transient grating techniques.

## Experimental set-up and Optical Components

A scheme of the set-up is shown in Figure 1. It is based on an amplified laser system delivering 8mJ 25fs pulses with a center wavelength of 800nm at 1kHz repetition rate. About 4.5mJ (peak power $2 \cdot 10^{11}$W) are used in the high harmonic generation (HHG), the remainder can serve for pump–probe experiments (not shown). The beam that drives the HHG is focused into the vacuum chamber by a lens with a curvature of
$r_c$ = 1030mm or $r_c$ = 763mm, depending on the experiment. These correspond to a focal length of f = 2270mm and f = 1680mm at 800nm, respectively. Care is taken that the beam diameter at the entrance port of the HHG chamber is sufficiently large to avoid beam profile deterioration due to nonlinear effects in the quartz window. The exact position of the focus can be altered by moving the lens.

The T-shaped HHG cell is machined from aluminum rod of varying length and contains a gas column of 4.4mm diameter. The cell ends are capped with 100μm thick copper or aluminum foil which are perforated by the laser. Hence the cell does not need to be precision aligned to the laser, but the foils are exchanged on a regular basis. The gas inlet is at the center of the cell where the backing pressure is controlled by a fine dosing valve. The backing pressure of the cell is about 10 torr. A turbo molecular pump with ~1600 liters/s pump speed keeps the vacuum around $10^{-3}$ torr in the chamber containing the gas cell leaking through the laser drilled holes. The cell also generates high harmonics if not sealed (at slightly higher pressures and roughly 5 times lower efficiency) which is convenient for alignment purposes. In that case, the cutoff harmonic is lower by about 10 eV under comparable parameters when operating an open gas cell.

After the HHG cell, the 800nm laser fundamental and EUV light propagate collinearly in vacuum. Separation of EUV light from the IR is achieved by a Si mirror[42,43] on which they are incident at an angle of 85 degrees to the surface normal, close to Brewster angle of 75 degrees for 800nm. Thereby a major amount of the ~$10^6$ times stronger fundamental is removed from the EUV beam path. A 150nm thick aluminum foil (purchased either from Luxel or Lebow) serves as filter for the IR light reflected off of the silicon mirror. Without the Si mirror, the Al filter would be destroyed by the intense fundamental beam.

The EUV light is refocused by a toroidal gold mirror (f = 650mm) with a tangiental radius of curvature of 9194mm and a saggital radius of curvature of 156.6mm, at which the light is incident at an angle of incidence AOI = 82.5 degrees to the surface normal. In studies of molecular dynamics, the sample sits at the focal spot of the toroidal mirror and the probe light is dispersed by a spectrometer (not shown). Figure 1 shows a spectrometer set-up in which the toroidal mirror is used in combination with a flat grating to characterize the HHG source. The gold grating (Edmund optics, 830lines/mm, blazing angle 19.38 degrees, AOI = 70 degrees) is placed downstream from the toroidal mirror to spectrally disperse the EUV light onto a microchannel plate (MCP) / phosphor screen detector and record on a charge coupled device (CCD). We rotate the grating to display different sections of the spectrum which heavily distorts the aspect of the spectra. Since the HHG polarization is identical to that of the IR, we need to operate the beamline with p polarized EUV light.

The reflectance of silicon and gold at the respective AOI and polarization have been calculated based on the data in Ref [44] and are represented in Figure 2. Since the EUV reflectivity drops rapidly when the AOI at the silicon mirror approaches the Brewster angle for 800nm, the AOI chosen is a compromise between suppressing the unwanted IR and extinguishing the EUV light. As can be seen in Figure 2, the reflectance of both, Au and Si, is higher for s-polarized EUV light especially in the photon energy range of a few tens of eV. Yet, only at p-polarization the Brewster condition can be exploited to suppress the IR.

The 20 to 70 eV transmittance window of an ideal 150nm Al aluminum foil shown in Figure 2 is valid in the absence of surface oxidation. The transmission is lowered by 25% for filters supported by a nickel mesh. Surface oxidation can be modeled as a 3nm thick $Al_2O_3$ layer on each surface which reduces the transmittance to 40 to 70% in the 30 to 70eV spectral range.[45] In practice, the losses on the filtering and refocusing optics are ca. 90% in the 20 to 70 eV range with steep increases on both the low and high energy side.

## Online Monitoring of HHG

The toroidal mirror has a reflectivity of 50-60% for p-polarized EUV radiation in the spectral region of interest and at the angle of incidence used in our set-up. The 40% of the light which is not reflected, is absorbed and partially cause emission of photoelectrons from the gold surface. We use the photocurrent to measure the overall yield and monitor the stability of HHG during experiments. To this end, we measure for each pulse the drain current flowing from the mirror to ground through an amplifier (SRS low noise charge sensitive amplifier model SR570 with a 6db 1kHz bandpass filter and a sensitivity of 2nA/V). This way of monitoring the HH radiation does not require adding elements to the beam path, and can be conveniently implemented in any set-up that uses an electrically conducting reflective element for EUV radiation. We use the device as online tool to access total HHG efficiency and stability without adding losses in the beam line.

Figure 3 shows the gold mirror, which is contacted at its edge via a copper foil. The left part of Figure 3 shows a representative reading from the oscilloscope. The integral over the positive part of the signal corresponds to the charge drawn for each laser pulse. The negative part stems from ringing of the amplifier. When interfaced with a computer, a single shot analysis of the HH flux is possible with this method. The noise is determined by the amplification process and can be estimated from the signal in the absence of EUV radiation. For the configuration in our lab, the error is on the order of 1%.

The yield of photoelectron ejection from a thick gold layer in the EUV region has been documented.[46] Assuming unity detection efficiency and taking a typical reading of $5 \cdot 10^{-13}$ Coulomb on the HH monitor tool, we calculate an integrated flux of $10^8$ photons per laser pulse in the whole spectrum impinging on the gold mirror. This yields a lower limit of $4 \cdot 10^{-7}$ for the conversion efficiency in the HHG cell for each harmonic in the 20 to 35 eV energy range when considering the reflectivity and transmission of the optics in the EUV beam path. In a pump-probe set-up, $10^7$ EUV photons arrive at the sample per laser pulse.

## High Harmonic Generation

A spectrum of the high harmonics is shown in Figure 4. The wavelength dependent reflectivity of the grating used to disperse the spectrum is not known so the spectrum is not corrected for this effect. A flat gold surface has an EUV reflectivity of 5% at this angle (that would correspond to the zero order of the grating), rather independent of the exact photon energy.

The harmonics are identified by their separation on the screen and the calculated dispersion angles. It is remarkable that the 11$^{th}$ harmonic at 77nm is present in the spectrum although the transmittance of the Al filter at this wavelength is low. This is due to the higher conversion efficiency towards lower harmonics, as they are closer to the perturbative domain of harmonic generation. At the high energy side, the spectrum is limited by the harmonic cutoff which is lower than expected due to an astigmatic beam profile at the HHG cell, resulting in lower intensity compared to a perfect Gaussian mode.

The yield of EUV radiation is determined by three important parameters: i) the length of the generation medium $L_{med}$; ii) the coherence length $L_{coh}$, i.e. the length over which the light generated by the individual emitters interferes constructively; and iii) the absorption length $L_{abs}$, i.e. the length over which the intensity of the EUV radiation decreases to $e^{-1}$ due to absorption by the medium.[47] The coherence as well as the absorption length are wavelength dependent.

The coherence length is determined by the phase mismatch between fundamental and high harmonics. Prominent factors for the mismatch are the intensity dependent dipole phase $\phi(I)$ of the harmonics as well as dispersion from the neutral medium and plasma and the so-called Gouy phase shift. Owing to the effect of $\phi(I)$ and the Gouy phase, a focus position before the generation medium enhances the HH yield.[24–26,28] Loose focusing or propagation inside a waveguide reduces the importance of $\phi(I)$ and the Gouy phase and dispersion and absorption become important.[48,49]

From a simplified one dimensional model, optimal yield is predicted for $L_{coh} > 5 \cdot L_{abs}$ and $L_{med} > 3 \cdot L_{abs}$.[48] Shorter media do not allow for a coherent build–up of intensity while for longer cell lengths absorption diminishes the conversion efficiency as the IR beam looses intensity with propagation. Even at the unphysical assumption of infinite coherence length, the yield saturates due to absorption. At the parameters given before, the efficiency reaches half of that asymptotic value.[48]

In practice, phase matching is accomplished by adjusting the length and density of the generation medium, the driving laser intensity, focal length and focus position such as to maximize the flux for the photon energy range of interest.[47,50]

Midorikawa and co-workers[50,51] as well as l'Huillier and co-workers[52] studied the influence of cell length on the HH yield in a loose focusing geometry and with cells of several centimeter length. In addition they examined the quality and divergence of the emitted HH beam. Systematic studies of HHG parameters in a single cell have also been reported by Ditmire et al.[28] using thin cells (<1 mm) and Seres et al.[35] using multiple cells.

We determined the influence of medium length for the f=2260mm lens. Figure 5 shows the overall yield of HHG in argon, as obtained from the HH monitor tool, for cells with a length varying from 6 to 47 mm. For the data summarized in Figure 5, all other parameters (gas pressure, laser pulse chirp, position of the focus with respect to the cell center) were optimized for overall HH yield for each cell.

As expected, the yield changes significantly with cell length and shows a maximum when it approaches half the Rayleigh length, which equals 52.6mm (28.8) for the lens with a focal length of f=2270mm (1680mm). There is a steep increase below this optimum length while the yield levels off at longer cell lengths.

For a pressure of 10 torr, the absorption length $L_{abs}$ in argon ranges from 1mm (H11) to 30mm (H29) for the harmonics present in the spectrum shown in Figure 4.[53] Comparing Figure 5 with a one dimensional model,[48] we can infer that the coherence length $L_{coh}$ in this setup is close to $5 \cdot L_{abs}$ since the experimental data reproduces both, the rise and slight decay of efficiency depicted in Figure 1 of Ref. [48] Indeed, $L_{abs}$ is reasonably close to 1/3 of the cell length for the dominant harmonics. Moreover, we find our results to be in line with those of Takahashi et al.[50] and similar to the parameters used by He et al.[52] or Wernet et al.[33]

# Outlook

Ultrafast time–resolved electronic spectroscopy in the EUV spectral region is capable of unprecedented temporal and spectral resolution and makes the complete dynamics of any molecular system traceable in photoelectron spectroscopy. The information gain will facilitate the interpretation of complex transient spectra. Electron correlation effects well known from steady–state techniques can be followed in real–time.

High harmonic generation can deliver ultrashort laser pulses of high coherence and low divergence in this spectral range with table-top commercial equipment. While many proof-of-principle experiments have already shown the applicability of HHG as EUV light source in ultrafast experiments, substantial improvements in photon flux and stability are necessary for a more routine use. Transient absorption is a photon hungry technique that requires probe light of high brightness and stability. Photoelectron detection does not need high pulse energies but benefits from high repetition rate sources. Naturally, a high photon flux is imperative for multiphoton and multidimensional spectroscopies, the latter of which are yet to be demonstrated at such short wavelengths.

Finally, the techniques are about to conquer after the gaseous and solid phase also the liquid phase which is most relevant for chemistry.[20,54]

## Acknowledgements

J. G. would like to acknowledge support by the European Research Agency via the FP-7 PEOPLE Program (Marie Curie Action 298210). M.K. acknowledges funding from the Austrian Science Fund (FWF, Erwin Schrödinger Fellowship, J 3299-N20). M. G. acknowledges funding via the Office of Science Early Career Research Program through the Office of Basic Energy Sciences, U.S. Department of Energy. This work was supported by the AMOS program within the Chemical Sciences, Geosciences, and Biosciences Division of the Office of Basic Energy Sciences, Office of Science, U.S. Department of Energy.

# Figures

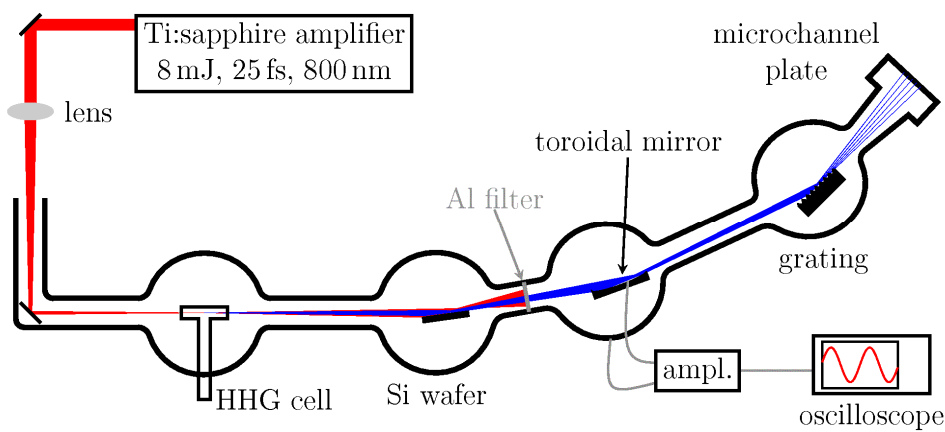

Figure 1: Experimental set-up. The 800nm laser beam is focused into the HHG cell with a long focal length, free to move lens, generating EUV light through HHG. The 800nm light is removed by a combination of Si wafer and Al filter. For this study the EUV light is spectrally dispersed onto the detector while for pump-probe studies, a sample (photoexcited by a fraction of the IR pulse) would reside at the focus of the toroidal mirror. The efficiency and stability of the EUV can be monitored on an oscilloscope.

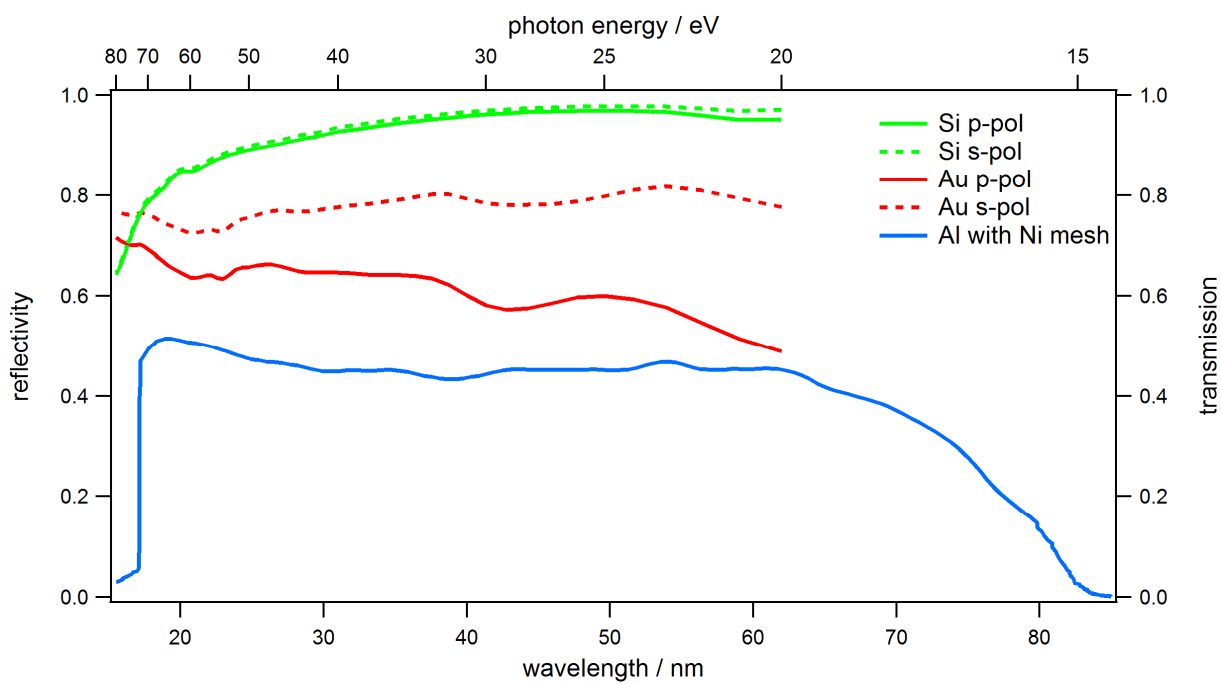

Figure 2: Calculated (see text) reflectivity of silicon (AOI = 85 degrees) and gold (AOI 82.5 degrees) and measured[55] transmittance of a 150nm thick aluminum foil (AOI = 0 degrees;).

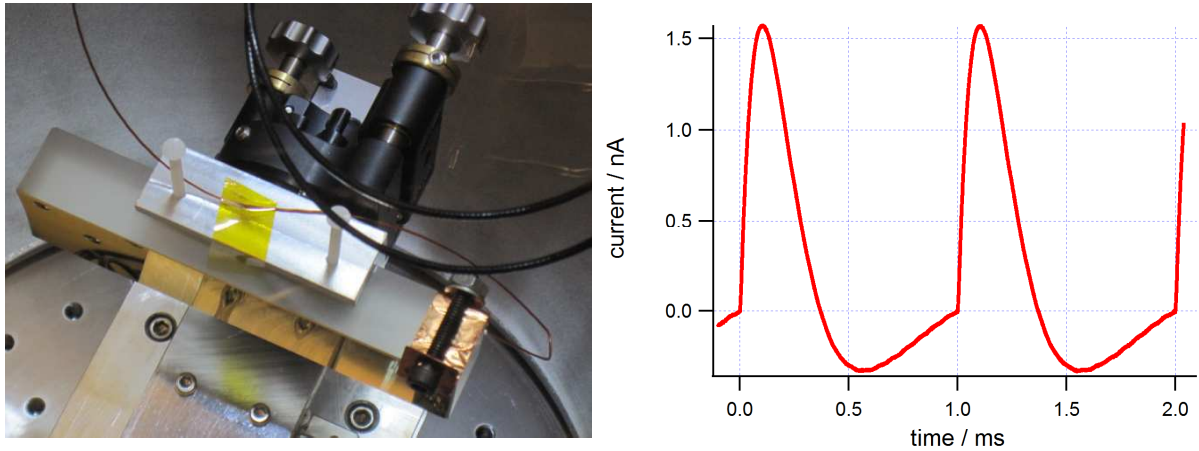

Figure 3: Left: Au mirror contacted by a copper sheet to measure the photocurrent. Right: High Harmonic Monitor reading on the oscilloscope. An integration over the scope trace results in the integrated charge from a single EUV pulse.

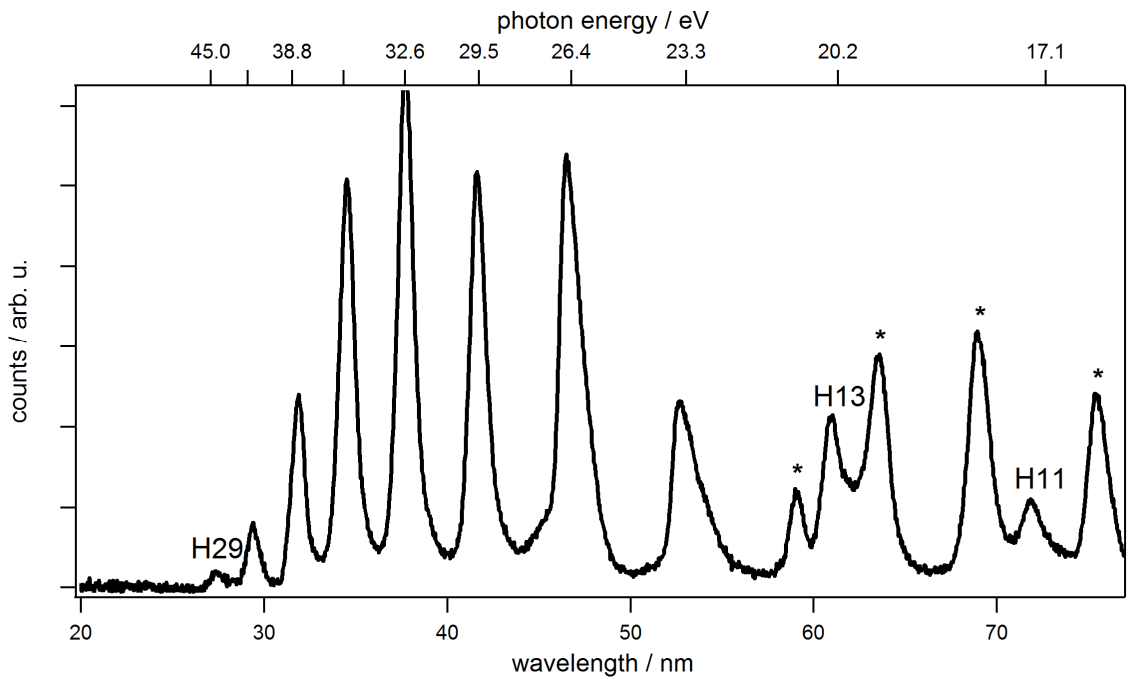

Figure 4: Spectrum of the high harmonics generated in a 60mm long cell with Ar at $I_{laser} = 1.3 \cdot 10^{14}$ W/cm$^2$ (f = 1680mm). The peaks marked with asterisks are assigned to 2$^{nd}$ order reflection of the grating. The spectrum is not corrected for the grating reflectivity and detector response.

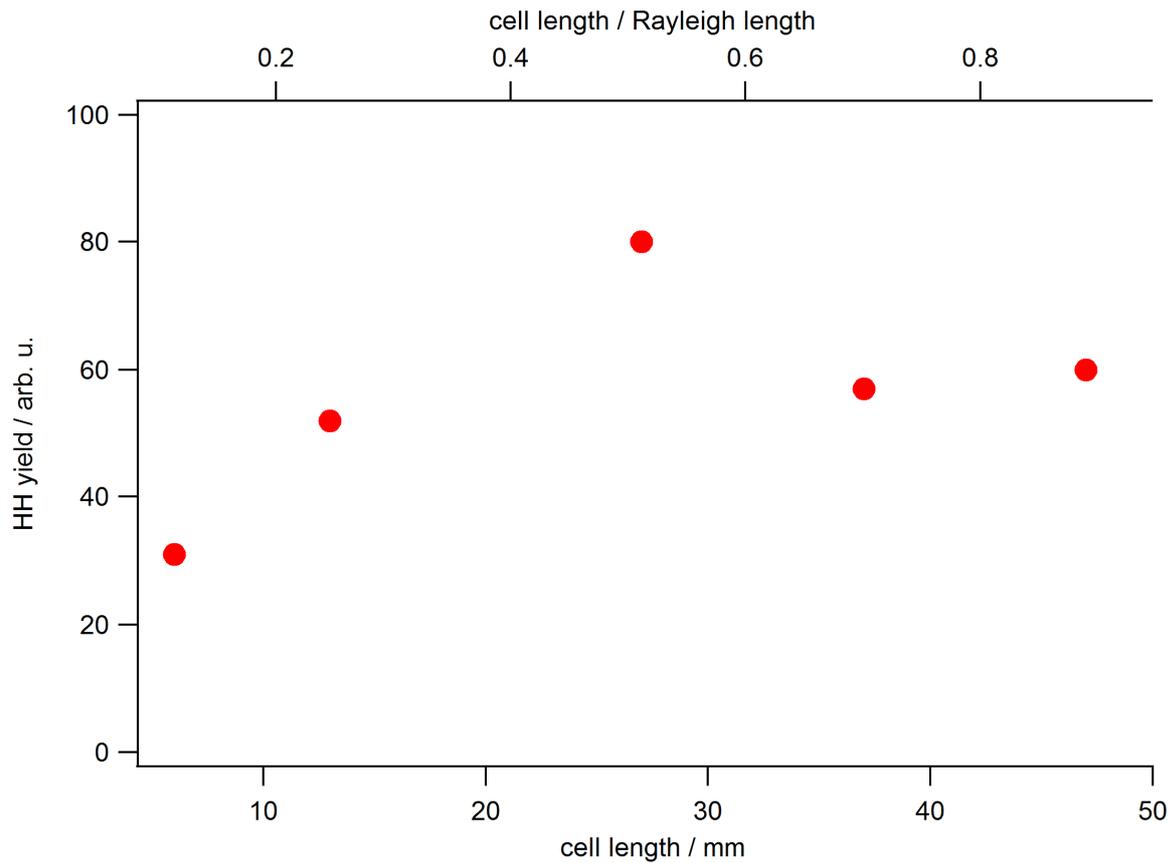

Figure 5: Efficiency of HHG in argon as a function of HHG cell length. $I_{laser}$ = 6·10$^{13}$W/cm$^2$, f = 2270mm. The rise and decay is in qualitative agreement with a theoretical model in Ref. [48] assuming an absorption length in the few mm to cm regime.